\documentclass[twocolumn,prb,amsmath,amssymb]{revtex4}

\usepackage{graphicx}
\usepackage{bm}

\begin{document}

\hyphenation{GaMnAs}
\hyphenation{GaMnP}
\hyphenation{GaMnN}
\hyphenation{GaMnSb}
\hyphenation{CdMnTe}
\hyphenation{CdMnSe}
\hyphenation{PbSnMnTe}

\title{
How to make semiconductors ferromagnetic: A first course on spintronics
}
\author{S. \surname{Das Sarma}}
\author{E. H. Hwang}
\author{A. Kaminski}

\affiliation{
Condensed Matter Theory Center,
Department of Physics, University of Maryland, College Park, Maryland
20742-4111
}

\begin{abstract}
The rapidly developing field of ferromagnetism in diluted magnetic
semiconductors, where a semiconductor host is magnetically doped by
transition metal impurities to produce a ferromagnetic semiconductor
(e.g. Ga$_{1-x}$Mn$_x$As with $x \approx 1-10\%$), is discussed with
the emphasis on elucidating the physical mechanisms underlying the
magnetic properties. Recent key developments are summarized with
critical discussions of the roles of disorder, localization, band
structure, defects, and the choice of materials in producing good
magnetic quality and high Curie temperature. The correlation between
magnetic and transport properties is argued to be a crucial ingredient
in developing a full understanding of the properties of ferromagnetic
semiconductors.
\end{abstract}

\maketitle

\section{introduction}

The study of semiconductors (i.e. intrinsically insulating materials
with band gaps in the $\sim$ 1 eV range) which are also ferromagnetic
has a long, but perhaps somewhat intermittent, history. Back in the
1960s and early 1970s there was a great deal of interest\cite{R1Eu} in
Eu-based chalcogenide materials which exhibited ferromagnetism with
Curie temperature around 50K or less. These were strongly insulating
systems with rather poor semiconducting transport properties. In the
1980s there were major efforts\cite{R2} in magnetically doped II--VI
ternary and quarternary semiconductor alloys, e.g. CdMnSe, CdMnTe,
PbSnMnTe \emph{etc.}, but most of these materials exhibited spin glass
or related disordered magnetic behavior and any ferromagnetism
invariably had very low ($\le$ 5K) transition temperature. The idea of
a diluted magnetic semiconductor, where a small ($< 10\%$)
concentration of magnetically active atoms (most often Mn although Co
and Cr have also been used occasionally) is distributed at the cation
sites of the host semiconductor, is appealing because such a system
may have both semiconducting and ferromagnetic properties.  Making
such materials has, however, been problematic because the magnetic
transition metal (i.e. Mn) is not thermodynamically stable in the
semiconductor host (e.g. GaAs), and tends to segregate.

The need for actively controlling and manipulating the carrier spin
degree of freedom (in addition to the charge degree of freedom) in
semiconductors leading to novel multifunctional
magneto-opto-electronic capabilities has created the emerging
interdisciplinary field of spintronics (or spin electronics). A key
component of spintronics is the development of new ferromagnetic
semiconductors. Following the successful development of
Ga$_{1-x}$Mn$_x$As \cite{one,two,two-bis} and In$_{1-x}$Mn$_x$As
\cite{three} as ferromagnetic semiconductors (with $x \approx 1-10\%$)
using careful low-temperature molecular beam epitaxy (MBE) technique,
intensive worldwide activity has led to claims of ferromagnetism (some
at room temperatures and above) in several magnetically doped
semiconductors, e.g. GaMnP,\cite{four} GaMnN,\cite{five}
GeMn,\cite{six} GaMnSb.\cite{seven} It is at present unclear whether
all these reports of ferromagnetism (particularly at room temperatures
or above) are indeed intrinsic magnetic behavior or are arising from
clustering and segregation effects associated with various
Mn-complexes (which have low solubility) and related materials
problems. The observed ferromagnetism of GaMnAs is, however,
well-established and is universally believed to be an intrinsic
diluted magnetic semiconductor (DMS) phenomenon.

The Mn dopants in GaMnAs serve the dual roles of magnetic impurities
providing the local magnetic moments and of acceptors producing, in
principle, one hole per Mn atom. The number density $n_{\textrm{c}}$
of the charge carriers (holes in GaMnAs), however, turns out to be by
almost an order of magnitude lower than the number density
$n_{\textrm{i}}$ of the Mn ions, making the system to be a highly
compensated doped semiconductor. This high level of compensation is
believed to be a crucial ingredient in the underlying DMS
ferromagnetic mechanism. In particular, the precise role played by the
relative values of the local moment density $n_{\textrm{i}}$ and the
hole density $n_{\textrm{c}}$ in giving rise to DMS ferromagnetism is
currently being debated in the literature to the extent that there is
no agreement even on whether the heavy compensation ($n_{\textrm{c}}
\ll n_{\textrm{i}}$) in the system helps or hinders ferromagnetism. An
important question in this context is to obtain the functional
dependence of the ferromagnetic transition temperature (``Curie
temperature'') $T_c$ on magnetic moment and hole densities
$n_{\textrm{i}}$ and $n_{\textrm{c}}$. In Ga$_{1-x}$Mn$_x$As the early
experimental results \cite{one,eight} indicated a nonmonotonic
behavior of $T_c(x)$ with $T_c$ first increasing with the Mn
concentration $x$, reaching a maximum of about 110K for $x \approx 5
\%$, and then decreasing with further increase of Mn concentration
above this optimum value. A strong correlation between the magnetic
transition temperature and transport properties of GaMnAs was also
reported in these findings \cite{one,eight} with the system exhibiting
relatively stronger metallic behavior close to this optimum Mn doping
level ($\sim 5\%$). Recent experimental studies \cite{nine,ten} of
Ga$_{1-x}$Mn$_x$As under carefully controlled annealing conditions
seem to suggest that the metastable nature (and high defect content)
of low temperature MBE grown GaMnAs may be playing an important role
in determining magnetic properties -- in particular, careful annealing
could, under suitable conditions, lead to an enhancement of $T_c$ with
increasing Mn content leading to an eventual saturation of the
transition temperature in the $x=5-8\%$ range without the nonmonotonic
$T_c(x)$ behavior observed earlier. These annealing studies
\cite{nine,ten} confirm the close connection between carrier transport
properties and ferromagnetism in DMS materials showing that typically
$T_c$ values are the highest in GaMnAs samples with nominally highest
conductivities (and presumably the highest hole densities). It has
therefore been suggested that, in agreement with the mean-field theory
(MFT) for DMS ferromagnetism originally derived in
Refs.~\onlinecite{eleven,c2} in a different context and recently
rediscovered in Refs.~\onlinecite{twelve,thirteen}, $T_c$ in GaMnAs
grows monotonically as $T_c \propto n_{\textrm{c}}^{1/3}$ when the
carrier density increases for a fixed Mn concentration. This simple
Weiss MFT \cite{eleven,c2,twelve,thirteen,fourteen} then implies a
convenient technique for enhancing the ferromagnetic transition
temperature in GaMnAs, namely, to increase the hole density as much as
possible. The general validity of this Weiss MFT-based assertion
remains questionable, and it has been argued\cite{c3,c3-bis} that this
prediction holds only in perturbative regime of small $J$ and small
$n_{\textrm{c}}$.

The GaMnAs annealing studies bring up important and interesting
questions regarding the nature and role of metastable defects in the
system that are presumably being healed during annealing leading to
$T_c$ enhancement. Two different kinds of GaMnAs lattice (atomistic)
defects have been discussed in this context. First, there are As
antisite (As$_{\textrm{Ga}}$) defects, As atoms sitting at Ga sites,
that are invariably present in the low temperature MBE growth required
for DMS GaMnAs. These As antisite defects act as double donors with
each As defect capturing two holes in GaMnAs. The second type of
defects that has been much discussed recently, is the Mn interstitial
defect, where the ``defective'' Mn atom sits at an interstitial site
in the GaAs lattice instead of occupying a substitutional cation site
replacing a Ga atom. The Mn interstitials are thought to be
particularly deleterious in affecting the magnetic properties of
GaMnAs because they have two adverse effects: first, they act as
donors (similar to As antisite defects) capturing holes; second, the
Mn interstitials typically have strong local direct antiferromagnetic
coupling to any neighboring substitutional Mn atoms, thereby
suppressing the net effective Mn magnetic moment available for DMS
ferromagnetism. The quantitative abundance of As antisites and Mn
interstitials in GaMnAs is difficult to estimate because there is
really no direct experimental technique to observe these defects. It
is, however, widely believed that the low-temperature MBE technique,
crucial for the homogeneous single-phase growth of GaMnAs, invariably
generates both of these defects with their relative importance varying
from sample to sample depending on the uncontrollable details of
growth and processing. In addition to the atomistic As antisite and
interstitial defects in GaMnAs, there are invariably many other
unintentional (as well as unknown and uncontrolled) defects and
impurities (both charged and neutral) in GaMnAs because low
temperature MBE, in general, produces materials of rather poor
quality. This poor quality of GaMnAs is reflected in the transport
properties as even the best GaMnAs samples seem to have very low
carrier mobilities. The measured hole mobilities in the so-called
``metallic'' GaMnAs systems are only about $1-50$ cm$^2$/V$\cdot$s
corresponding to a typical carrier mean free path of $0.1-1$ nm. Such
small values of mean free paths would indicate a strong tendency
toward carrier localization in GaMnAs, and indeed there are reports of
reentrant metal-insulator transitions in GaMnAs as a function of the
Mn content $x$. Specifically, it has been reported\cite{eight} that
Ga$_{1-x}$Mn$_x$As is an insulator for small ($x < 3 \%$) and large
($x > 7\%$) values of $x$ whereas it is metallic (albeit with very
small mean free paths) in the intermediate range of $x$ ($\sim 5 \%$)
where the ferromagnetic transition temperature is maximized. Other
experimental studies (often using careful sample annealing procedures)
do not find such reentrant metal-insulator transition
behavior.\cite{nine,ten}

It is important to emphasize that the existence of DMS ferromagnetism
seems to be independent of the system being metallic or insulating.
For GaMnAs, both metallic and insulating samples are ferromagnetic
with the transition temperature typically being higher for metallic
systems although this may not necessarily be a generic behavior. Many
other systems exhibiting DMS ferromagnetism, e.g., InMnAs, GeMn,
GaMnSb, GaMnN, are however strongly insulating, and as
such, metallicity (i.e. itinerant nature of the carriers), is not a
necessary pre-condition for the existence of ferromagnetism in DMS
systems.  Additionally, in Ga$_{1-x}$Mn$_x$As samples exhibiting
reentrant metal-insulator transitions the ferromagnetic behavior
appears to be completely continuous as a function of $x$ with the only
observable effect being a variation in $T_c$, which is expected since
different GaMnAs samples, even with nominally the same value of $x$,
show wide variations in $T_c$ values.  The real DMS carriers mediating
the ferromagnetic interaction between the local moments are
generically likely to be far from free holes in the valence band of
the host semiconductor material. They are, in all likelihood, extended
or bound carriers in the impurity (or the defect) band which forms in
the presence of the Mn dopants. There is a great deal of direct
experimental support for the relevance of this impurity band picture
for DMS ferromagnetism, and first principles band structure
calculations also confirm the impurity band nature of the carriers
active in DMS ferromagnetism.  Both the impurity band nature of the
carrier system and the presence of strong charge and spin disorder in
the system (due to the random Mn dopants as well as other defects,
disorder, and impurities invariably present in the DMS material) make
carrier localization a relevant issue in DMS ferromagnetism. We
believe that understanding the correlation between transport and
magnetic properties is a key ingredient in DMS physics.

In the next section of this review we critically discuss some of the
proposed theoretical models (and their validity or applicability) for
DMS ferromagnetism emphasizing the connection between magnetization
(as well as the Curie temperature) and transport properties. We
conclude in section III with a brief discussion of (many) open
questions and some possible future directions in the subject.

This article is is more of a theoretical perspective; it is not meant
to be an exhaustive review of the vast literature on ferromagnetic
semiconductors. We only discuss some of the key developments in the
subject in order to highlight the physical mechanisms and the
important system parameters controlling ferromagnetism in diluted
magnetic semiconductors.

\section{Theories}

The fact that the DMS are ferromagnetic independent of their weakly
metallic or strongly insulating nature implies a robust character for
the underlying mechanism leading to the long range magnetic order in
these systems. Clearly the ferromagnetic mechanism should not depend
crucially on the carrier system being ``free'' valence band holes
since the strongly insulating DMS systems do not have any free holes.
The currently accepted picture for DMS ferromagnetism is that it is
the local antiferromagnetic coupling between the carriers (i.e., holes
in GaMnAs) and the Mn magnetic moments that leads to long range
ferromagnetic ordering of Mn local moments. The carrier system also
becomes spin-polarized in the process with the carrier magnetic moment
directed against the Mn magnetic ordering by virtue of the
antiferromagnetic hole-Mn coupling, but the total magnetic moment of
the spin polarized carriers is extremely small since $n_{\textrm{c}}
\ll n_{\textrm{i}}$ and $|{\bf S}| > |{\bf s}|$ where ${\bf S}$ and
${\bf s}$ are respectively the Mn and the hole spin. The relevant DMS
effective magnetic Hamiltonian can be written as
\begin{equation}
H_M = \int d^3rJ({\bf r}) {\bf S}({\bf r}) {\bf \cdot} {\bf s}({\bf
  r}),
\label{H_M}
\end{equation}
where ${\bf S}({\bf r})$ and ${\bf s}({\bf r})$ are respectively the Mn and
hole spin densities. The coupling $J({\bf r})$ between Mn local
moments and hole spins can, in principle, be ferromagnetic ($J<0$) or
antiferromagnetic ($J>0$), but the effective interaction between the
Mn local moments mediated by the holes (through $H_M$) is always
ferromagnetic. The magnitude of $J$ must come from a first principles
band structure calculation or from experiments.

Since the Mn moments are localized at specific lattice sites, it is
sensible to write Eq.~(\ref{H_M}) in the more conventional form
\begin{equation}
H_M = J \sum_i \int d^3 r \delta({\bf r}-{\bf R}_i)
{\bf S}_i {\bf \cdot} {\bf s}({\bf r})\;,
\label{H_Mp}
\end{equation}
where the sum over $i$ goes over the Mn sites in the GaAs lattice, and
we have assumed the exchange coupling $J$ to be a constant. The $J{\bf
  S}_i {\bf \cdot} {\bf s}$ local exchange coupling defined in
Eq.~(\ref{H_Mp}) is sometimes referred to as the Zener model (or the
$s-d$ coupling although in the context GaMnAs, where the carriers are
holes, the coupling is more like an $p-d$ exchange coupling between Mn
$d$ levels and the $p$ valence band).

The magnetic Hamiltonian defined in Eqs.~(\ref{H_M}) and (\ref{H_Mp})
is, of course, one of the most well-known interaction Hamiltonians in
all of condensed matter physics. Depending on the relative magnitudes
of hole and Mn moment densities $n_{\textrm{c}}$ and $n_{\textrm{i}}$,
and the coupling strength $J$, $H_M$ describes the Kondo model or the
Kondo lattice model or $s-f$ (or $s-d$) Zener model or the double
exchange model or the RKKY model or (in the presence of strong
disorder) the (RKKY) spin glass model. Most of the earlier studies of
this exchange Hamiltonian have concentrated on the low
impurity-density limit, $n_{\textrm{i}} \ll n_{\textrm{c}}$, where the
local moments are a perturbation on the carrier Fermi surface.  The
DMS systems provide an interesting novel regime for studying $H_M$,
namely the regime of low carrier density $n_{\textrm{i}} \gg
n_{\textrm{c}}$. There has not been much earlier work focused on the
low carrier-density limit of the Kondo lattice Hamiltonian or the RKKY
spin glass Hamiltonian, which is the appropriate parameter regime for
GaMnAs. We emphasize that Kondo physics plays no essential role in DMS
systems, but because of the strong inherent disorder in these systems
spin glass physics may very well be playing a role at lower values of
$n_{\textrm{i}}/n_{\textrm{c}}$.

The simplest way to understand DMS ferromagnetism on a qualitative
level is to neglect all disorder effects, and assume that the system
can be thought of as a collection of local moments of density
$n_{\textrm{i}}$ interacting with itinerant holes of density
$n_{\textrm{c}}$. In the weak coupling regime ($J \ll t, E_F$ where
$t$ and $E_F$ are the band width and the Fermi energy of the carrier
system) it is then possible to ``eliminate'' the carrier degrees of
freedom by mapping the problem into the corresponding RKKY problem of
Mn local moments interacting indirectly via the holes through $H_M$.
This carrier-polarization-mediated RKKY exchange interaction is
essentially a second order perturbation theoretic calculation in $J$,
leading to the following interaction Hamiltonian between the Mn spins
(assuming, for the sake of simplicity, a single parabolic hole band)
\begin{equation}
H_i = \sum_{i\ne j}J_{ij} {\bf S}_i {\bf \cdot} {\bf S}_j,
\label{H_i}
\end{equation}
with
\begin{equation}
J_{ij} \propto |J|^2 \left [ \frac{\sin(2k_F R) -(2k_F R) \cos(2k_F
  R)}{(2k_FR)^4} \right ], 
\label{J_ij}
\end{equation}
where $k_F \propto n_{\textrm{c}}^{1/3} \propto E_F^{1/2}$ is the
Fermi wave vector for the holes and $R \equiv |{\bf R}_{ij}| \equiv
|{\bf R}_i - {\bf R}_j|$ is the distance between the Mn impurities at
$i$ and $j$ lattice sites. This oscillatory RKKY indirect exchange
magnetic coupling between local moments placed in an electron gas
arises from the spin polarization induced in the electron system by
the local moment, and is a perturbative effect of the local moments on
the carrier system (valid in the limit of small $J$). In the DMS
systems an approximate average distance between the Mn local moments
is $\bar{R} \sim n_{\textrm{i}}^{1/3}$, making typical values of $k_FR
\sim (n_{\textrm{c}}/n_{\textrm{i}})^{1/3}$. For ferromagnetism to
occur (remembering that the real GaMnAs system is intrinsically
strongly substitutionally disordered) one must avoid the strong
frustration effects associated with random sign changes in the
magnetic interaction of Eq.~(\ref{H_i}) arising from the oscillatory
terms in Eq.~(\ref{J_ij}), which can only be guaranteed when $k_FR \ll
1$. This then necessitates the condition
$n_{\textrm{c}}/n_{\textrm{i}} \ll 1$, making strong compensation,
i.e., $n_{\textrm{c}} \ll n_{\textrm{i}}$, a necessary condition for
ferromagnetism in the system.

We note that a Weiss MFT calculation can easily be
carried out for the coupled Mn-hole magnetic problem defined by the
Zener-RKKY model of Eqs.~(\ref{H_M})-(\ref{J_ij}). For a simple
single-band degenerate hole model, the result for $T_c^{\textrm{MF}}$ in the
Weiss MFT was obtained a very long time ago:\cite{eleven,c2}
\begin{equation}
T_c^{\textrm{MF}} = \frac{S(S+1)}{3} s^2 |J|^2 D(E_F)n_{\textrm{i}},
\label{T_c}
\end{equation}
where $S$ is the local moment spin and $s$ is the carrier spin. The
carrier density of states $D(E_F)$ in Eq.~(\ref{T_c}), in the
degenerate one band free carrier system, is given by $D(E_F) \propto
E_F^{1/2} \propto n_{\textrm{c}}^{1/3}$. Thus $T_c^{\textrm{MF}}$ increases
monotonically with the exchange coupling, the local moment density,
and the carrier density as
\begin{equation}
T_c^{\textrm{MF}} \propto J^2 n_{\textrm{c}}^{1/3}n_{\textrm{i}},
\label{T_cp}
\end{equation}
where $n_{\textrm{c}}$ and $n_{\textrm{i}}$ are treated as independent
density parameters.  Based on this simple Weiss MFT (and its various
extensions incorporating spin wave corrections and/or explicit
spin-orbit coupling effects in the GaAs valence band), quantitative
predictions have been made\cite{fourteen} about the dependence of the
ferromagnetic transition temperature in various DMS systems on the
system parameters (carrier and local moment densities, effective
exchange coupling strength, etc.).  In addition, we have recently
shown\cite{sixteen} that the MFT for the model defined by Eqs.
(\ref{H_M}) and (\ref{H_Mp}) can be extended to the nondegenerate
carrier system more appropriate for the localized holes in the
strongly insulating DMS materials. In contrast to the degenerate
results of Eq.  (\ref{T_cp}), which is appropriate for the metallic
(``extended'') carrier system (e.g. free valence band holes), the
nondegenerate MFT provides a transition temperature dependence given
by
\begin{equation}
T_c^{\textrm{MF}} \propto |J|(n_{\textrm{c}}n_{\textrm{i}})^{1/2}.
\label{T_cpp}
\end{equation}
Thus, the dependence of $T_c^{\textrm{MF}}$ on the system parameters changes
drastically in the localized nondegenerate limit. More details can be
found in Ref. \onlinecite{sixteen}.

The simple Weiss mean-field theory (Eqs. (\ref{T_c}) and (\ref{T_cp}))
based on the model of degenerate valence band holes interacting with
the Mn local moments (and its obvious extensions) has been used
extensively in the literature. The question, therefore, arises about
its validity, or more generally, about its quantitative reliability.
It is argued that the molecular MFT should apply well in DMS systems
because of the large length scales (e.g. $k_F^{-1}$) involved in the
problem. We believe that there are good reasons to question the
validity of the Weiss MFT as applied to DMS ferromagnetism.  First,
strong disorder in the system invariably brings in some effects of
frustration which should become more significant at larger values of
carrier density, making the simple MFT defined by Eqs. (\ref{T_c}) and
(\ref{T_cp}) increasingly incorrect at higher carrier densities (even
for the metallic system). In fact, $T_c$ should vanish in the
disordered system as $n_{\textrm{c}} \rightarrow n_{\textrm{i}}$.
Second, MFT in this problem being equivalent to a second order
perturbation theory in the exchange coupling $J$ (i.e. Eqs.
(\ref{H_M})--(\ref{T_cp})), the validity of Weiss mean-field
predictions becomes highly questionable for large values of $J$ where
carriers may bind to individual local moment sites to lower their
energy forming an impurity band. There is significant experimental
evidence for impurity band formation in DMS materials, which is not
included in the MFT.  Therefore mean-field predictions will fail at
larger values of $J$ where DMS magnetic properties should be dominated
by impurity band physics not caught by Eqs. (\ref{T_c}) and
(\ref{T_cp}). We have recently developed a dynamical mean-field theory
(DMFT) picture\cite{c3,c3-bis} of DMS ferromagnetism where impurity
band physics is explicitly included. Third, in the limit of
$n_{\textrm{i}} \gg n_{\textrm{c}}$, where DMS ferromagnetism
prevails, disorder effects are extremely strong (as is obvious from
the insulating nature of most DMS materials and from the extremely low
mobilities of the so-called ``metallic'' DMS materials which may very
well also be insulating at $T=0$), and therefore the degenerate
``metallic'' Weiss MFT approach based on the carriers being the free
extended valence band holes of the host semiconductor may be
fundamentally flawed. A more appropriate picture in this disordered
localized regime may be the percolation picture of DMS ferromagnetism
recently developed by us\cite{c4} where each hole is bound in a
cluster of magnetic impurities forming a bound magnetic polaron. With
decreasing temperature, these bound magnetic polarons coalesce
eventually leading to a magnetic percolation transition at $T_c$ where
a magnetic percolation cluster spans the whole system.  This polaron
percolation picture of DMS ferromagnetism is strongly
supported\cite{Dagotto} by direct numerical simulations and by
experimental studies showing strongly concave magnetization curves in
many DMS systems which follow naturally from our
theory.\cite{sixteen,c4}

Below we provide some details on the dynamical mean-field and polaron
percolation theories of DMS ferromagnetism.

\subsection{Dynamical mean-field theory}

In this section we briefly describe the ``dynamical mean-field
theory'' (DMFT). Due to size limitations, we are not able to give
detailed description of this formalism here. A reader interested in
more details may find them in Ref.~\onlinecite{c3,c3-bis,sixteen}.

We model the Ga$_{1-x}$Mn$_x$As system as a lattice of sites, which
are randomly nonmagnetic (with probability $1-x$) or magnetic (with
probability $x$), where $x$ now indicates the relative concentration
({\it i.e.} per Ga site) of active Mn local moments in
Ga$_{1-x}$Mn$_x$As. The DMFT approximation amounts to assuming that
the self energy is local or momentum independent, $\Sigma({\bf
  p},i\omega_n) \rightarrow \Sigma(i\omega_n)$, and then all of the
relevant physics may be determined from the local
(momentum-integrated) Green function defined by
\begin{equation}
G_{\textrm{loc}}(i\omega_n) = a_0^3 \int \frac{d^3p}{(2\pi)^3} \frac{1}{
i\omega_n +\mu - \epsilon(p) 
-\Sigma_{\sigma}(i\omega_n)},
\end{equation}
where we have normalized the momentum integral to the volume of the
unit cell $a_0^3$, and $\mu$ is the chemical potential.
$G_{\textrm{loc}}$ is in general a matrix in spin and band indices and
depends on whether one is considering a magnetic or non-magnetic
 site. Since $G_{\textrm{loc}}$ is a local function, it is the
solution of a local problem specified by a mean-field function $g_0$,
which is related to the partition function $Z_{\textrm{loc}}=\int
d\textbf{S} \exp(-S_{\textrm{loc}})$ with action
\begin{eqnarray}
S_{\textrm{loc}}&=& \sum_{\alpha \beta }g_{0\alpha \beta }^{a}(\tau
-\tau ^{\prime 
})c_{\alpha }^{\dagger}(\tau 
)c_{\beta }^{\vphantom{\dagger}}(\tau ^{\prime })\nonumber\\
&+&J \sum_{\alpha \beta }\left(\textbf{S}\cdot\bm{\sigma}_{\alpha
  \beta}\right)
c_{\alpha}^{\dagger}(\tau)c_{\beta }^{\vphantom{\dagger}}(\tau')\;, 
\end{eqnarray}
on a magnetic site and 
\begin{equation}
S_{\textrm{loc}}=\sum_{\alpha \beta } g_{0\alpha \beta }^{b}(\tau
-\tau ^{\prime }) 
c_{\alpha }^{\dagger}(\tau)c_{\beta }^{\vphantom{\dagger}}(\tau ^{\prime }),
\end{equation}
on a non-magnetic site.
Here $c_{\alpha}^{\vphantom{\dagger}}(\tau)$
($c_{\alpha}^{\dagger}(\tau)$) is the destruction (creation) operator of a
fermion in the spin state $\alpha$ and at time $\tau$.
$g_0(\tau - \tau')$ plays the role of the Weiss mean field (bare
Green function for the local effective action $S_{\textrm{loc}}$) and is
a function of time.
The magnetic-site mean-field function $g_{0}^{a}$ can be written as
$g_{0\alpha \beta }^{a}=a_{0}+a_{1}{\textbf{m}}\cdot
\bm{\sigma}_{\alpha \beta }$ with ${\textbf{m}}$ the
magnetization direction 
and $a_{1}$ vanishing in the paramagnetic state ($T>T_c$). 
Since the spin axis is chosen parallel to {\bf m}, $g_0^a$ becomes a
diagonal matrix with components parallel ($g_{0\uparrow}^a = a_0 +
a_1$) and antiparallel ($g_{0\downarrow}^a = a_0 - a_1$) to {\bf m}.
It is specified by
the condition that the local Green function computed from
$Z_{\textrm{loc}}$, namely $\delta \ln Z_{\textrm{loc}}/\delta
g_{0}^{a}=\left( g_{0}^{a}-\Sigma \right) ^{-1}$ is identical to the
local Green function computed by performing the momentum integral
using the same self energy.

\begin{figure}
\includegraphics[width=3.25in]{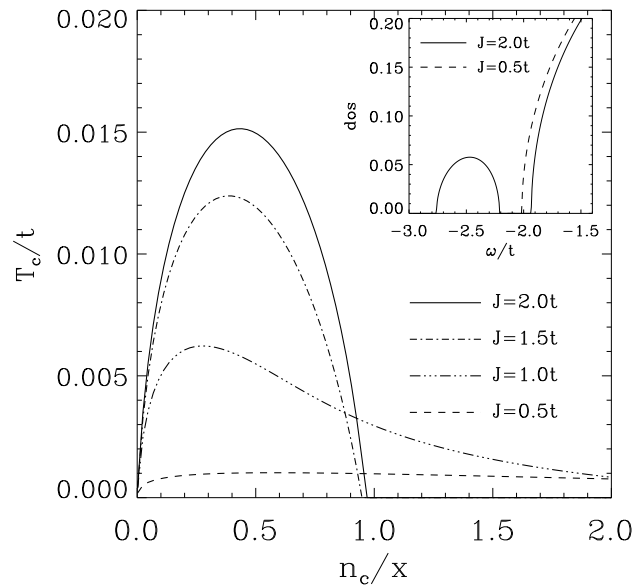}
\caption{Calculated  ferromagnetic transition temperature $T_c$ as a
  function of carrier carrier density $n_{\textrm{c}}/x$ for $x=0.05$
  and various exchange couplings $J$. Inset shows the majority spin
  density of states at $T=T_c$ and for $x=0.05$ and $J= 0.5t,\ 1.0t$.
}
\label{fig_tc}
\end{figure}

Within this approximation we calculate the normalized magnetization of
the local moments  with the semicircle density of states, $D(\omega) =
\sqrt{4t^2-\omega^2}/2\pi t$, where $t$ is a bandwidth of the system.
The normalized magnetization $M(T)$ is given by
$M(T) = \int (d{\bf S}) \cdot \hat {\bf S}
\exp(-S_{\textrm{loc}})/Z_{\textrm{loc}}$. 
As the temperature is increased, the spins disorder and eventually the
magnetic transition temperature is reached. Above this temperature,
$g_{0}$ is spin-independent.  By linearizing the equation in the
magnetic part of $g_0$ with respect to $a_1$ we may obtain the
ferromagnetic transition temperature $T_c$.  The details on the
calculation of $T_c$ are given in Refs.~\onlinecite{c3,c3-bis}.  
In Fig. \ref{fig_tc} we show the calculated $T_c$ as a function of
carrier density for $x=0.05$ and various couplings $J$, and 
the corresponding density of states are given in the inset. When $J < t$
the impurity band formation dose not occur (see the inset for
$J=0.5t$) and $T_c$ are saturated as the density increases.  
However, as $J>t$ the impurity spin bands are well established (see
the inset for $J=2.0t$). In this limit the $T_c$ has a maximum value
around the half filling of the impurity band ($n_c \approx 0.5x$) and
$T_c$ vanishes as the carriers fill the band ($n_c=x$) because no low
energy hopping processes are allowed in a ferromagnetic state for a
filled impurity band. Within DMFT the calculated $T_c$ shows the
nonmomotonic behavior for all $J$ and the Weiss mean field prediction
$T_c \propto n^{1/3}$ can be applied only for small $n_c$.

\begin{figure*}
\includegraphics[width=6.5in]{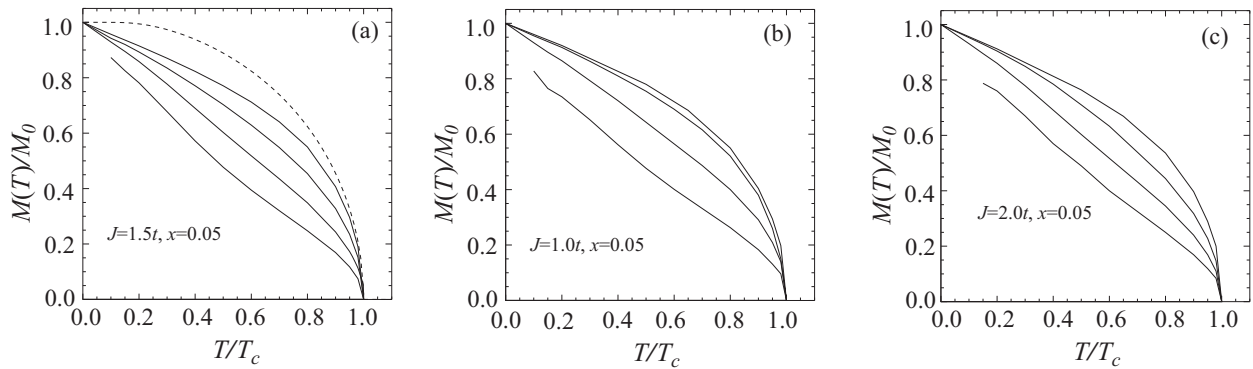}
\caption{
\label{DMFT_fig8}
The normalized DMFT impurity magnetization 
as a function of temperature for (a) $J=1.5t$, (b) $J=1.0t$, and (c)
$J=2.0t$, for $x=0.05$ and for  
$n_{\textrm{c}}/n_{\textrm{i}}=$0.4, 0.2, 0.1, 0.04 (from
the top). 
The dashed line in (a) represents the magnetization calculated 
for the simple Weiss mean field theory for the local moment spin $S=5/2$.
}
\end{figure*}

In Fig.~\ref{DMFT_fig8} we show the normalized magnetization of the
local moments as a function of temperature for different values of
$J=$1.0, 1.5, 2.0$t$ and $x=0.05$, and for various hole densities,
$n_{\textrm{c}}/n_{\textrm{i}}$.  For the small coupling constant
$J=t$ the impurity band is not formed, but for $J=$1.5, 2.0$t$ we have
a spin polarized impurity band.  For relatively high density
($n_{\textrm{c}}/n_{\textrm{i}}=0.4$) the calculated
magnetization looks similar to the Weiss mean-field results.  But for
low density ($n_{\textrm{c}}/n_{\textrm{i}}=0.04$) we have a linear
$M(T)$ in the intermediate temperature range.  Near the critical
temperature $T_c$ the critical behavior of the magnetization for all
density is given by $M(T) \propto (T_c-T)^{1/2}$. For different
exchange couplings we have similar results (\emph{i.\ e.}, linear
behavior at low densities and intermediate temperature ranges).

In particular, it should be possible to obtain the $M(T)$ behavior for
the localized carrier case also from DMFT by incorporating impurity
band localization in the DMFT formalism. Our current theory does not
include localization, and the impurity band (or valence band) carriers
in our DMFT calculations are all delocalized metallic carriers. First
principles band theory calculations indicate that the actual exchange
coupling in Ga$_{1-x}$Mn$_x$As may be close to critical $J_c$, and as
such impurity band physics may be quite important for understanding
DMS magnetization. To incorporate physics of localization one needs to
include disorder effects (invariably present in real DMS systems) in
the model. All our DMFT calculations are done in the virtual crystal
approximation where effective field is averaged appropriately leaving
out random disorder effects explicitly. In the next section we
explicitly incorporate disorder in the theory by developing a
percolation theory approach to DMS magnetization for the strongly
localized insulating systems.

\subsection{Percolation theory}

Our percolation theory for DMS\cite{c4} applies strictly
in the regime of strongly localized holes where the dynamical
mean-field theory for delocalized carriers described above has little
validity.  These two theories, mean-field theory and percolation
theory, are therefore complementary.  

The percolation theory assumes the same model of carrier-mediated
ferromagnetism, but now the carriers are pinned down with the
localization radius $a_B$. The disorder, averaged out in the
mean-field theory, plays a key role in the carrier localization.  The
magnetic impurities are assumed to be completely randomly distributed
in the host semiconductor lattice in contrast to the mean-field case
where the carrier states are free and the disorder is neglected.

As we have demonstrated in Ref.~\onlinecite{c4}, the problem of
ferromagnetic transition in a system of bound magnetic polarons can be
mapped onto the problem of overlapping spheres well-known in the
percolation theory.\cite{EfrosShklovskiiBook} The latter problem
studies spheres of the same radius $r$ randomly placed in space
(three-dimensional in our case) with some concentration $n$.
Overlapping spheres make ``clusters;'' as the sphere radius $r$
becomes larger, more and more spheres join into clusters, the clusters
coalesce, and finally, at some critical value of the sphere radius, an
infinite cluster spanning the whole sample appears.  This problem has
only one dimensionless parameter, $r^3n$, and therefore can be easily
studied by means of Monte-Carlo simulations.

\begin{figure}
\includegraphics[width=3in]{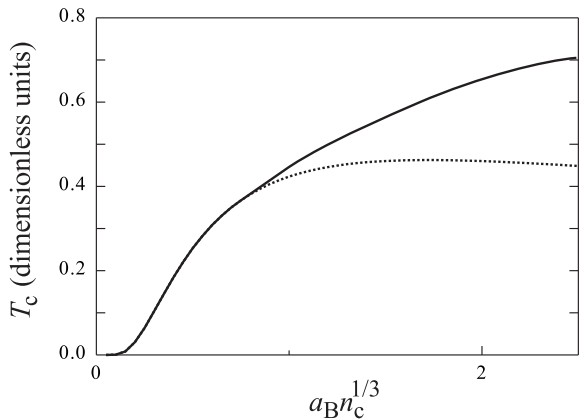}
\caption{\label{fig:Tc_vs_aBnh} Curie temperature $T_c$ as a function of the 
  dimensionless parameter $a_B^3 n_{\textrm{c}}$. At $a_B^3
  n_{\textrm{c}}\lesssim 1$, $T_c$ is given by Eq.~(\protect\ref{Tc}).
  At $a_B^3 n_{\textrm{c}}\gtrsim 1$, Eq.~(\protect\ref{Tc}) (being
  beyond limits of its applicability ) predicts decline of $T_c$
  (dotted line); in reality, $T_c$ grows monotonically with $a_B^3
  n_{\textrm{c}}$ (solid line), though its exact behavior is unknown.
}
\end{figure}

Each sphere of the overlapping spheres problem corresponds to a bound
magnetic polaron, which is a complex formed by one localized hole and
many magnetic impurities with their spins polarized by the exchange
interaction with the hole spin. The concentration $n$ of spheres
coincides with the concentration $n_{\textrm{c}}$ of localized holes. The
expression for the effective polaron radius in terms of the physical
parameters of the system under consideration is not trivial and has
been found in our earlier work:\cite{c4}
\begin{equation}
\label{params}
r^3n = \left[0.86+\left(a_B^3 n_{\textrm{c}}\right)^{\frac{1}{3}}
\ln \frac{T_c}{T}\right]^3\;.
\end{equation}
Here $0.64\approx 0.86^3$ is the critical value of the parameter
$r^3n$ at which the infinite cluster appears, and $T_c$ is the Curie
temperature of the ferromagnetic system under consideration, derived
in Ref.~\onlinecite{c4},
\begin{equation}
\label{Tc}
T_c\sim sSJ\left(\frac{a_0}{a_B}\right)^3
\left(a_B^3n_{\textrm{c}}\right)^\frac{1}{3}\!
\sqrt{\frac{n_{\textrm{i}}}{n_{\textrm{c}}}}\ 
\exp\left(-\frac{0.86}{\left(a_B^3n_{\textrm{c}}\right)^\frac{1}{3}}\right)\;.
\end{equation}
The limit of applicability of Eq.~(\ref{Tc}) is determined by the
condition $a_B^3 n_{\textrm{c}} \ll 1$. The dependence of $T_c$ on
$a_B^3 n_{\textrm{c}}$ is shown in Fig.~\ref{fig:Tc_vs_aBnh}.

\begin{figure}
\includegraphics{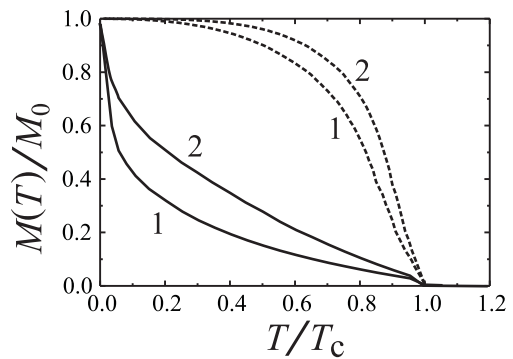}
\caption{\label{fig:graphs} The solid lines show the
  relative magnetization of the magnetic impurities
  [Eq.~(\protect\ref{magnfinal})] for $a_B^3 n_{\textrm{c}} = 10^{-3}$
  (curve 1) and $10^{-2}$ (curve 2). The dashed lines show the
  relative magnetization of localized holes, whose contribution to the
  total sample magnetization is small. }
\end{figure}

Since $n_{\textrm{c}} \ll n_{\textrm{i}}$, one polaron includes many
magnetic impurities, and the total magnetization of the sample is that
of impurities:
\begin{equation}
\label{magnfinal}
\frac{M(T)}{M_0}= 
{\cal V}\left(0.86+\left(a_B^3 n_{\textrm{c}}\right)^{\frac{1}{3}}
\ln \frac{T_c}{T}\right),
\end{equation}
where the universal function ${\cal V}(y)$ is the infinite cluster's
volume in the model of overlapping spheres; it depends only on the
product $y$ of the spheres' diameter and the cubic root of their
concentration.  One can see that the shape of the magnetization curve
is determined by only one dimensionless parameter $a_B^3
n_{\textrm{c}}$, while the expression (\ref{Tc}) for $T_c$ is more
complicated and depends on all parameters of the model.
Fig.~\ref{fig:graphs} shows the temperature dependence of the
magnetization at two values of $a_B^3 n_{\textrm{c}}$; the curve is
more concave at smaller values of this parameter.  The concave shape
of the curve is consistent with the experimental magnetization data in
Ge$_{1-x}$Mn$_x$\cite{six} and low-$T_c$ III-V
samples\cite{three,seven} where our polaron
percolation picture applies better due to stronger carrier
localization associated with lower values of $a_B^3 n_{\textrm{c}}$.
Our magnetization results also agree with the numerical results of
Ref.~\onlinecite{sds10,Dagotto}.  Very recently, a direct numerical
Hartree-Fock calculation has convincingly verified the magnetic
polaron percolation picture for the localized regime.\cite{c6}

\section{Conclusion}

When carriers are present with a carrier density $n_c$ in a
semiconductor containing a certain magnetic impurity density
$n_{\textrm{i}}$ of local magnetic moments, a Weiss mean field theory,
developed years ago,\cite{eleven,c2} within the Zener-RKKY effective
magnetic Hamiltonian describing the local exchange coupling between
local moments and carrier spin, would always predict ferromagnetism
with a mean field transition temperature $T_c^{\textrm{MF}}$ varying linearly
in $n_{\textrm{i}}$ and $n_c^{1/3}$.  This Weiss mean field theory,
appropriately modified by the complicated spin-orbit coupled multiband
valence band structure of GaAs, has been extensively used in making
quantitative predictions about the magnetic properties of GaMnAs. This
Weiss mean field theory, while being a powerful qualitative tool, is
obviously too simplistic to be taken seriously for quantitative
analyses of DMS properties.  In real DMS systems spatial fluctuations
(i.e.  disorder associated with random positions of the magnetic
dopants) and thermal fluctuations of magnetic moments as well as
impurity band and discrete lattice effects (all neglected in the Weiss
mean field theory) are likely to be of real importance even if the
magnetic coupling can be assumed to be a simple local exchange
coupling between local impurity moments and carrier spins.  In
addition, the mean field theory averages out the oscillatory exchange
coupling between the impurity moments, which must be important at
higher carrier densities, particularly in the presence of strong
spatial disorder.  For example, the mean field theory incorrectly
predicts the ground state to be ferromagnetic for \emph{all} values of
$n_c$ and $n_{\textrm{i}}$, whereas in reality of course the system is
a spin glass for larger values of $n_c$.  The mean field theory is
qualitatively valid only for very small values of the
carrier-local-moment exchange coupling $J$ where impurity band effects
are weak, and applies only in the limit of $n_c\ll n_{\textrm{i}}$
where disorder and oscillatory aspects of exchange can be neglected.
In addition, the standard mean field theory is valid only for
degenerate metallic systems although a nondegenerate localized version
of the mean field theory has recently been developed.

There have been recent theoretical
developments\cite{c3,c3-bis,c4,sixteen} going beyond the simple Weiss
mean field theory where impurity band effects (both for
metallic\cite{c3,c3-bis} and insulating\cite{c4} DMS systems) are
explicitly included in the dynamical mean
field\cite{c3,c3-bis,sixteen} and magnetic polaron percolation
theories\cite{c4,sixteen}.  The DMFT theory predicts, in contradiction
with the Weiss mean field theory, that the ferromagnetic transition
temperature $T_c$ is not monotonic in the carrier density (and the
coupling strength $J$), and may vanish for larger values of $n_c$.
While the definitive quantitative theory for DMS ferromagnetism,
taking into account complete band structure complications of the
system along with disorder, fluctuation, and nonperturbative
strong-coupling (i.e. $J$ not necessarily small) effects in a
consistent manner is still lacking, it is increasingly clear that
disorder and impurity band aspects of the real DMS systems are
extremely important in understanding DMS magnetic properties, and
theories \cite{c3,c3-bis,c4,sixteen} that include these effects should
be of qualitative and quantitative significance in explaining DMS
magnetic properties.

The observed correlation between transport and magnetic properties in
DMS systems, e.g. the more metallic systems typically having more
convex magnetization curves as a function of temperature whereas more
insulating systems having extreme non-mean-field like concave
magnetization versus temperature curves (with considerable missing
magnetic moments), is perhaps a strong indicator of the underlying
mechanism of DMS ferromagnetism.\cite{sixteen} This correlation
indicates a carrier-mediated magnetic interaction origin of DMS
ferromagnetism.  In particular, two very recent theoretical
publications\cite{sixteen,c6} have discussed in considerable details
this correlation between metallicity and DMS ferromagnetism.  In
Ref.~\onlinecite{sixteen} it has been argued that the appropriate
picture for DMS ferromagnetism in strongly disordered insulating
materials is the bound magnetic polaron percolation picture whereas at
higher densities (and/or lower disorder), where the system is
metallic, the dynamical mean-field picture (which becomes equivalent
to the Weiss mean field picture at low values of exchange coupling
insufficient to cause the impurity band formation) is more
appropriate.  In Ref.~\onlinecite{c6} a detailed numerical simulation
of the disordered DMS model has been carried out qualitatively
validating this picture.  The question therefore arises regarding the
nature of the DMS ground state: Is it localized and insulating or
extended and metallic or is there a transition between the two as a
function of carrier density?  If there is a true metal-insulator
transition, then the percolation picture holds in the insulating phase
and the dynamical mean field picture holds in the metallic phase.
Although there have been experimental claims of metal-insulator
transitions in GaMnAs systems as a function of Mn content, the current
experimental situation is sufficiently murky that is possible that all
GaMnAs systems are in fact insulating at $T=0$ for all values of the
Mn content.  It would therefore be desirable to develop a dynamical
mean field theory (or even just the Weiss mean field theory) of DMS
ferromagnetism explicitly incorporating disorder effects so that the
magnetic properties can be studied continuously through the
metal-insulator transition (or crossover, as the case may be).

This work is supported by US-ONR and DARPA.

\bibliography{ssc}

\end{document}